THEME ARTICLE: Remote Learning and Work

# Empowering Database Learning through Remote Educational Escape Rooms

Enrique Barra, Universidad Politécnica de Madrid, Madrid, 28040, Spain

Sonsoles López-Pernas, University of Eastern Finland, Joensuu, 80100, Finland

Aldo Gordillo, Universidad Politécnica de Madrid, Madrid, 28040, Spain

Alejandro Pozo, Universidad Politécnica de Madrid, Madrid, 28040, Spain

Andres Muñoz-Arcentales, Universidad Politécnica de Madrid, Madrid, 28040, Spain

Javier Conde, Universidad Politécnica de Madrid, Madrid, 28040, Spain

**Abstract**—Learning databases is indispensable for individuals studying software engineering, computer science, or involved in the IT industry. We have analyzed a remote educational escape room for teaching databases in four different higher education courses in two consecutive academic years. We employed three instruments for evaluation: a pre- and post-test to assess the escape room's learning effectiveness, a questionnaire to gather students' perceptions, and a web platform that unobtrusively records students' interactions and performance. We show novel evidence that educational escape rooms conducted remotely can be engaging as well as effective for teaching databases.

Remote learning has become an increasingly vital component of education, training, and professional development. The COVID-19 pandemic has accelerated the adoption of remote learning worldwide while highlighting the importance of effective remote learning methods. Educators and trainers have been compelled to adapt their teaching techniques to cater to remote learners' needs, leading to the development of new and innovative approaches to instruction. In this article, we explore one such approach which is the use of remote educational escape rooms, specifically applied to the teaching of one important aspect related to software engineering: databases.

Nicholson described escape rooms as "live-action team-based games where players discover clues, solve puzzles, and accomplish tasks in one or more rooms in order to accomplish a specific goal (usually escaping from the room) in a limited amount of time"[1]. Escape rooms have become immensely popular in recent years due to their unique and engaging experience. They offer participants an interactive and immersive form of entertainment that challenges their problem-solving skills and encourages teamwork. The application of escape room activities to the educational context gave birth to educational escape rooms. Thus, an educational escape room is one designed specifically for learning purposes, incorporating learning materials into puzzles in a way that requires students to master these materials to successfully complete the escape room.

THEME/FEATURE/DEPARTMENT

The available body of literature presents compelling evidence that incorporating educational escape rooms into a curriculum can have a notably beneficial effect on student engagement and learning[2-6]. These activities have been thoroughly applied when teaching different topics within engineering and computer science, such as programming, software modeling, cryptography, and computer networks, but prior research is scarce when applying them to teaching databases. Databases are a critical component of many applications, and a fundamental understanding of them is essential for anyone studying software engineering, computer science, or working in the tech industry. Teaching databases combines important theoretical concepts with practical skills related to operating database technologies, importing and exporting data, querying, and even developing applications that work with data stored in databases and interact with them. This subject constitutes a suitable context for the application of educational escape rooms, to consolidate theoretical concepts, and to apply practical skills.

In this article, we examine the benefits of using an educational escape room conducted remotely to teach databases in four different higher education courses in two consecutive academic years. We analyze students' learning gains, perceptions, and performance when participating in the experiences.

## RELATED WORK

Educational escape rooms have been mostly conducted as face-to-face activities. However, several of them had to be transformed or even re-designed to be conducted remotely in response to the COVID-19 pandemic[7,8]. Conducting them as remote activities can be challenging, due to the difficulty of creating all the puzzles in a digital form and the possibility that the students get stuck at any stage and need additional help, which can make the escape room a stressful experience[9]. Nevertheless, conducting educational escape rooms in online distance learning settings can also pose multiple advantages. For instance, the activity can be scaled to a larger number of students as it is not restricted by the size of physical spaces, and it is easier to conduct it asynchronously.

In the five literature reviews conducted to date[2-6], more than 180 different educational escape room experiences have been analyzed. These experiences were applied to very diverse fields and topics, ranging from healthcare (e.g., pharmacology, toxicology, oncology, dentistry) to the field of technology (e.g., cybersecurity, programming, robotics, computer networks) or humanities (e.g., literature, history, philosophy). However, to the knowledge of the authors, only one study has been reported in the literature applying this kind of activity to teaching databases. The escape room was called "Aces of databases"[10] and was conducted in a classroom setting in a higher education course about Data Science at Universitat de Valencia in Spain. A total of 75 students participated, grouped into 21 teams, and a short survey with 8 questions was conducted after the activity to capture the student perceptions. In general, it was a successful activity that the students liked, but the teaching staff reported it was too laborious. The authors did not report on the learning effectiveness of the escape room or the performance of the students during or after the execution of the activity. In view of the shortcomings of the current body of knowledge, it can be concluded that further research on remote educational escape rooms for teaching databases is needed.

## DESCRIPTION OF THE ESCAPE ROOM EXPERIENCE

### Context

The present study evaluates an educational escape room conducted remotely in four different higher education courses in two consecutive academic years, 2021/2022 and 2022/2023. These courses were:

➔ "Non-Relational and Distributed Databases" course of the Bachelor of Science in Data Engineering and Systems, a second-year course that accounts for 6 ECTS (European Credit Transfer System) credits, corresponding to 150-180 hours of student work.

➔ "Databases" course of the Bachelor of Biomedical Engineering. This course is a third-year course that accounts for 6 ECTS credits (150-180 hours).

➔ "Big Data: Foundations and Infrastructure" course of the Master's Degree in Telecommunication Engineering. This course is a second-year course that accounts for 6 ECTS credits (150-180 hours).

➔ "Information Systems and Databases" course of the Master's Degree in Network and Telematic Services Engineering. This master lasts one year and the course accounts for 3 ECTS credits (75-90 hours).

The objective of the educational escape room was to review and strengthen some of the fundamental concepts taught during the courses by providing a motivating and captivating experience. Students could participate remotely in the activity for extra credit. Students who chose to take part in the escape room earned 0.5 points to be added to the final grade out of 10, albeit with the condition of passing the exam to get the extra points.

## Design

The escape room activity was designed to be completed by the students within a two-hour timeframe at the time of their convenience in the last two weeks before the final exam. It was conducted online through a web platform called Escapp[11] that students could access from home, from school, or from elsewhere. This platform served various purposes, including student enrollment (in pairs or alone), content and multimedia resource management during the activity, verification of puzzle solutions, monitoring progress, managing hints, showing the teams the remaining time and the leaderboard, and handling student grading and attendance. The puzzles within the escape room were organized sequentially, ensuring that the solution to each puzzle unlocked the next one. A quiz-based hint approach was employed to assist students when they faced difficulties. To obtain a hint, teams needed to correctly answer at least four out of five random questions from a quiz about databases. Figure 1 shows a screenshot of a student computer where the Escapp platform can be seen with instructions and a video, a terminal with the database shell to run queries, and a map showing a crocodile-shaped hotel required to obtain the username of the wanted criminal.

## Narrative and puzzles

The "Ask Why" approach[12] was taken into account when elaborating the narrative of the game. This approach suggests that each puzzle, task, item, document, or resource existing in the escape room had a narrative-consistent reason for being there. The narrative reveals students´ role in the story, the problem that arose, and the final goal, together with some other guidance or intermediate goals. This narrative also justifies why the final goal should be achieved before the deadline.

The overall concept of the narrative was "Investigate a Crime or Mystery" which, according to Nicholson's survey[1], is a theme employed in 9% of escape rooms worldwide. The experience started by presenting the students with a video where a cyberintelligence agent explains that he has been tracking a criminal for a long time due to multiple thefts of large amounts of money. The criminal plans to transform the money into untraceable cryptocurrencies in two hours and escape as his identity is unknown. Fortunately, the agent has managed to hack the criminals' password manager data, but it is in JSON format (a format used in document databases like MongoDB) and he needs help to deal with these data. That is the reason why he has contacted the top experts in NoSQL database technologies, the students of the course.

Figure 2 provides an overview of the sequential puzzles of the educational escape room. Each step represents an action students had to perform to solve the puzzles.

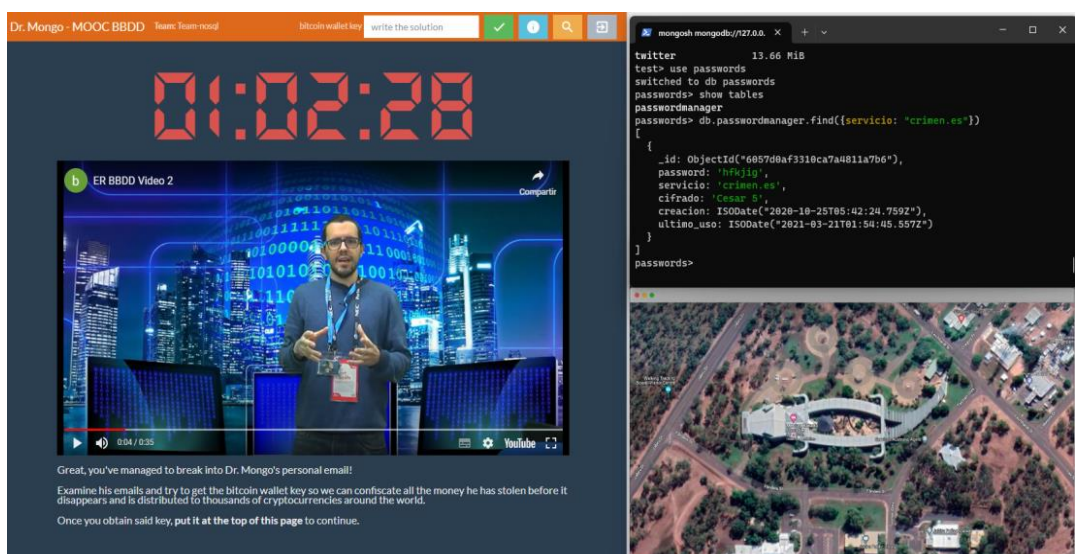

FIGURE 1. Screenshot of the Escapp platform, a terminal with a query and a map with a crocodile-shaped hotel.

THEME/FEATURE/DEPARTMENT

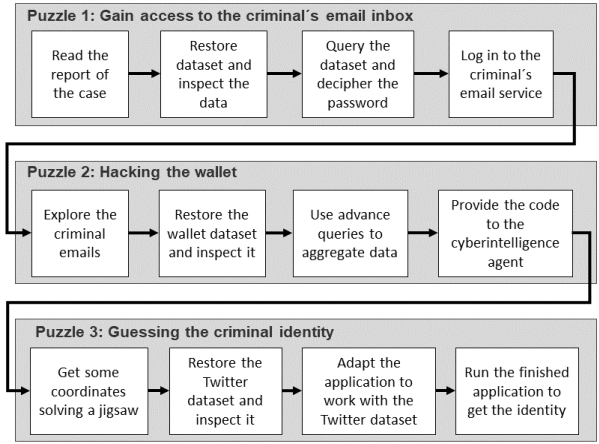

FIGURE 2. Escape room puzzles overview

# EVALUATION METHODOLOGY

In order to assess the impact of the remote educational escape room on the instruction of databases, we have examined its learning effectiveness, the perceptions of the students, and their performance during the activity. Our evaluation incorporated three instruments: (1) a pre- and post-test to gauge students' increase in knowledge, (2) a questionnaire to collect students' opinions, and (3) the Escapp web platform, which yielded data on students' actions while playing.

To accurately assess the learning effectiveness a pre-test was conducted right before the activity and a post-test afterwards. Students were not provided with feedback or the correct answers after the pre-test, preventing them from memorizing the answers. The questions were designed to evaluate the same knowledge and skills that were needed to solve the puzzles, and encompassed all the learning objectives addressed in the activity. Each test had the same 10 questions and a time limit of 10 minutes, and the results did not count toward students' final grades to avoid cheating.

Following the completion of the post-test, students were asked to participate in a survey to gather their opinions on the activity. The survey included closed-ended questions that focused on students' overall opinions and acceptance of the activity. Additionally, students were presented with a series of statements and asked to rate their level of agreement or disagreement using a 5-point Likert scale. These statements aimed to evaluate students' perceived learning effectiveness and attitudes towards the educational escape room as a learning method. Finally, students were invited to share their thoughts on various aspects of the escape room, including its design (difficulty, puzzle types, duration, hint approach), organization, and immersion.

Lastly, the Escapp web platform served as a comprehensive tool for capturing important information about student participation and performance in the escape room activity. It automatically recorded the number of puzzles solved by each team, the time taken to solve each puzzle, the hints requested and obtained by each team, and the teams that successfully completed the remote educational escape room, along with their total completion time.

# RESULTS AND DISCUSSION

A total of 299 students participated in the escape room: 144 men (48.2%), 150 women (50.1%), and 5 (1.7%) did not indicate gender, being 21.4 the mean age and 2.3 the standard deviation. 61,9% of the participants had participated in a ludic escape room before, but only 33.8% had previously participated in an educational one. When asked about the difficulty of the database course, practically half of the surveyed students (49.2%) believed that the subject was neither difficult nor easy, nearly a quarter of the students (25.7%) said the subject was easy for them, around another quarter expressed the opposite (25.1%).

Data have been aggregated among the different editions of the escape room since no statistical difference was perceived in the learning effectiveness of the activity according to a Kruskal-Wallis test. The average score for the pre-test was 5.8 (MED = 6.0, SD = 2.1) on a scale of 1 to 10, whereas the average score for the post-test was 7.24 (MED = 7.3, SD = 2.0). The average increase in scores was 1.4 (MED = 1.3, SD = 2.1).

To assess the difference between pre-test and post-test scores, the Wilcoxon Signed-Ranks Test for paired samples was employed. The magnitude of this difference was measured using the biserial correlation coefficient (r). Based on Cohen's guidelines[13], an r value between 0.1 and 0.3 indicates a small effect size, between 0.3 and 0.5 indicates a medium effect size, and an r value greater than 0.5 indicates a large effect size.

According to the results obtained in the Wilcoxon test, the difference between the post-test and pre-test scores was statistically significant with a large effect size (p-value < 0.001; r=0.69).

Table 1 shows the results of the questionnaire administered to collect students' perceptions toward the remote educational escape room. For each item, the mean (M), and standard deviation (SD) are shown.

The results of the survey show that students had a very positive overall opinion of the remote educational escape room. They stated that the experience was fun and immersive and slightly disagreed that it was a stressful experience. Overall, these results show that the remote educational escape room was a very compelling and interesting activity for the students. Further evidence of this fact was that almost all students stated that they would recommend other students to participate (97.6%) and that they would like other courses to implement similar activities (95.6%).

Regarding the learning effectiveness perceived by the students, results show that they believed that the activity allowed them to improve their knowledge, which is a finding consistent with the results obtained in the learning effectiveness evaluation with the pre-test and post-test.

Two computer lab sessions are conducted during the course, where the students go with their laptops and work on the practical exercises and assignments in a more guided way because they can ask questions or ask for help to the teaching staff. When comparing the escape room activity with computer lab sessions, students strongly agreed that they liked the former better, but only slightly agreed that they learned more. This is a very good result, as computer lab sessions are also highly valued activities in database courses, according to feedback received in recent years.

Concerning the design, organization, and difficulty of the activity, students thought they were prepared enough to complete the activity and neither agreed nor disagreed that it was too hard. The duration was also considered adequate. The hints approach was well valued, the hints received were rated as useful to progress in the escape room and students disagreed with the statement that they would have liked more help. These results confirm the appropriateness of the hint approach and its usefulness, which is a critical aspect in educational escape rooms, especially in remote ones. Students also agreed that the initial guidance provided was enough, an aspect also very important in remote asynchronous activities, and that in general, the activity was well organized.

Finally, the survey asked some questions related to the escape room being virtual and remote. Students had no problem communicating and collaborating with their teammates remotely and liked very much the fact that the escape room included digital puzzles. This comes as no surprise since nowadays higher education students (especially those in STEM programs) are fully accustomed to digital and collaboration tools. Students highly valued the possibility of participating autonomously at a convenient date and time for them and showed no preference for the escape room to have been held in person.

TABLE 1. Results of the student survey (N= 247).

| Question | M | SD |
|---|---|---|
| What is your general opinion on the escape room? (1 Very poor - 5 Very good) | 4.3 | 0.8 |
| Please, state your level of agreement with the following statements (1 Strongly disagree - 5 Strongly agree): | | |
| The escape room helped me improve my knowledge of the course materials | 4.1 | 0.8 |
| I liked the escape room to a larger extent than a laboratory session | 4.3 | 1.0 |
| I learned more with the escape room than I would have learned in a laboratory session | 3.6 | 1.2 |
| I believe I had enough preparation to succeed in the escape room | 3.9 | 1.0 |
| The escape room was exceedingly hard | 2.9 | 0.9 |
| The escape room was enjoyable | 4.2 | 0.9 |
| The escape room was immersive | 4.2 | 0.9 |
| The escape room was stressful | 2.8 | 1.3 |
| The escape room had an appropriate duration | 3.9 | 1.1 |
| The escape room employed an adequate hint approach | 4.1 | 1.0 |
| The hints provided were useful to progress in the activity | 3.8 | 1.2 |
| I would have liked to receive more help during the escape room | 2.6 | 1.2 |
| The initial instructions provided were enough | 3.9 | 1.0 |
| The escape room was well organized | 4.3 | 0.9 |
| I liked that the escape room contained digital puzzles | 4.3 | 0.8 |
| I liked to participate in the escape room independently without the need to adjust to a specific date and time | 4.5 | 0.9 |
| The escape room being conducted remotely created communication or collaboration problems in my team | 1.7 | 1.2 |
| I would have preferred the escape room to be carried out in person | 2.8 | 1.4 |
|  | Yes (%) | No (%) |
| Do you recommend other students to take part in the escape room? | 97.6 | 2.4 |
| Would you like other courses to integrate activities like the escape room? | 95.6 | 4.4 |

THEME/FEATURE/DEPARTMENT

In total 185 teams participated in the educational escape room, of which 114 (61.6%) were pairs and 71 (38.4%) were students participating alone. Table 2 shows the students´ performance in the escape room. 89 teams solved the last puzzle and so succeeded in the escape room, the mean time to solve the complete escape room for these teams was 93.4 minutes (SD = 17.3). We consider that the activity's difficulty as a whole was well-balanced, because almost half of the teams succeeded before the time ran out with some margin for the rest of the teams to continue trying to solve the puzzles and getting hints, and so learning. The number of hints obtained and failed to obtain show that the last puzzle was the most complicated one where the students needed more help as it included developing code for an application and not only doing queries.

TABLE 2. Students´ performance in the escape room

| Puzzle | Teams that solved each puzzle | Time invested | | Hints obtained | | Hints failed to obtain | |
|---|---|---|---|---|---|---|---|
| | | M | SD | M | SD | M | SD |
| 1 | 178 | 28.1 | 15.0 | 1.92 | 2.8 | 0.7 | 2.2 |
| 2 | 173 | 35.7 | 14.0 | 1.18 | 2.1 | 0.4 | 1.2 |
| 3 | 89 | 40.5 | 17.4 | 3.44 | 3.3 | 1.0 | 1.9 |

## CONCLUSIONS

Overall, we provide evidence that educational escape rooms conducted remotely pose a great opportunity to be included in the curriculum of database courses. These kind of activities are engaging, motivating, immersive, and fun for the students, and of course they are learning activities where students increase their knowledge about databases.

Conducting a virtual educational escape room remotely poses multiple advantages. It allows the participation of a large number of students without the need to book additional facilities. It provides significant flexibility to the teaching staff, who do not need to reserve one or multiple sessions to conduct the experience, like it would happen in face-to-face settings, but rather it can be complemented with lectures and lab sessions. Additionally it also provides great flexibility to the students, as they can participate in the experience at their preferred timeslot and when they feel they are prepared, they can even participate multiple times if they want to re-try the experience. Finally, if the activity is conducted in an unsupervised way (as was our case) it lessens the burden on the teaching staff compared to the same supervised activity.

A recent meta-analysis[14] has found that remote educational escape rooms are less effective than their face-to-face counterparts. However, existing research has not provided evidence on how remote educational escape rooms compare to other remote learning activities. Our results on the effectiveness of remote escape rooms for learning databases are promising, but further research is required to determine if students' knowledge acquisition is at least comparable to other remote learning approaches.

Creating the escape room in the first instance is complex and requires a good design and integration with the rest of the course materials and activities. The investment of time and effort by the course staff to design and create educational escape rooms is notably higher than that of other traditional hands-on learning activities such as computer lab sessions. According to our estimates, creating the educational escape room took teachers between 30 and 40 hours of effort, while an equivalent lab session takes between 8 and 12 hours. However, the significant and undeniable positive effect on student engagement and learning makes it a worthwhile endeavor. This is accentuated by the fact that the escape room is easily reusable in another course on the same topics or in successive editions of the same course and this way many students can benefit from its advantages.

Further research can include the application of these activities to MOOC courses or other remote learning settings to reduce student dropout rates and increase engagement.

## ACKNOWLEDGMENTS

The authors would like to acknowledge the support of the FUN4DATE (PID2022-136684OB-C22) project funded by the Spanish Agencia Estatal de Investigación (AEI) 10.13039/501100011033.

**Enrique Barra** is an Associate Professor at Universidad Politécnica de Madrid. His research interests include the generation and distribution of educational content, games and social networks in education. He received his PhD in Engineering from Universidad Politécnica de Madrid. Contact him at enrique.barra@upm.es

**Sonsoles López-Pernas** is a Senior Researcher at University of Eastern Finland. Her research interests include educational escape rooms and learning analytics. She received her PhD in Engineering from Universidad Politécnica de Madrid. Contact her at sonsoles.lopez@uef.fi.

**Aldo Gordillo** is an Associate Professor at Universidad Politécnica de Madrid. His research interests include game-based learning and educational technology. He received his PhD in Engineering from Universidad Politécnica de Madrid. Contact him at a.gordillo@upm.es.

**Alejandro Pozo** is an Assistant Professor at Universidad Politécnica de Madrid. His research interests include data engineering and cybersecurity. He received his PhD in Engineering from Universidad Politécnica de Madrid. Contact him at alejandro.pozo@upm.es.

**Andres Muñoz-Arcentales** is an Assistant Professor at Universidad Politécnica de Madrid. His research interests include big data and data engineering. He received his PhD in Engineering from Universidad Politécnica de Madrid. Contact him at joseandres.munoz@upm.es.

**Javier Conde** is an Assistant Professor at Universidad Politécnica de Madrid. His research interests include open data and digital twins. He received his PhD in Engineering from Universidad Politécnica de Madrid. Contact him at javier.conde.diaz@upm.es.